\pdfoutput=1

\documentclass[entropy,article,accept,moreauthors,pdftex]{Definitions/mdpi} 

\firstpage{1} 
\pubvolume{23}
\issuenum{11}
\articlenumber{1506}
\pubyear{2021}
\copyrightyear{2021}
\externaleditor{Academic Editors: Sebastian Deffner, Raymond Laflamme, Juan Pablo Paz and Michael Zwolak}
\datereceived{8 October 2021} 
\dateaccepted{9 November 2021} 
\datepublished{13 November 2021} 
\hreflink{https://doi.org/\linebreak 10.3390/e23111506}

\usepackage{array}
\usepackage{mathtools}
\usepackage{amssymb}
\usepackage{braket}
\usepackage{bbm}

\allowdisplaybreaks

\makeatletter

\theoremstyle{definition}

\theoremstyle{remark}

\theoremstyle{plain}

\theoremstyle{definition}

\newcommand{\tr}{\operatorname{tr}}

\renewcommand\bra[1]{{\langle{#1}|}}
\makeatletter
\renewcommand\ket[1]{%
  \@ifnextchar\bra{\k@t{#1}\!}{\k@t{#1}}%
}
\newcommand\k@t[1]{{|{#1}\rangle}}
\makeatother

\usepackage{bm}% bold math

\providecommand{\definitionname}{Definition}
\providecommand{\examplename}{Example}
\providecommand{\remarkname}{Remark}
\providecommand{\theoremname}{Theorem}

\Title{Thermality Versus Objectivity: Can They Peacefully Coexist?}

\TitleCitation{Thermality Versus Objectivity: Can They Peacefully Coexist?~}

\Author{Thao P. Le
 $^{1,}$*\orcidA{}, Andreas Winter $^{2,3}$\orcidB{} and Gerardo Adesso $^{1}$\orcidC{}}
\AuthorNames{Thao P. Le, Andreas Winter and Gerardo Adesso}

\AuthorCitation{Le, T.P.; Winter, A.; Adesso, G.}

\address{%
$^{1}$ \quad School of Mathematical Sciences, University of Nottingham, University Park, Nottingham NG7 2RD, UK; gerardo.adesso@nottingham.ac.uk\\
$^{2}$ \quad Instituci\'{o} Catalana de Recerca i Estudis Avan\c{c}ats (ICREA), Pg. Lluis Companys, 23, 08001 Barcelona, Spain; andreas.winter@uab.cat\\
$^{3}$ \quad Grup d'Informaci\'{o} Qu\`{a}ntica, Departament de F\'{i}sica, Universitat Aut\`{o}noma de Barcelona, 08193~Bellaterra (Barcelona), Spain}

\corres{Correspondence: thao.le.16@ucl.ac.uk}

\abstract{Under the influence of external environments, quantum systems can undergo various different processes, including decoherence and equilibration. We observe that macroscopic objects are both objective and thermal, thus leading to the expectation that both objectivity and thermalisation can peacefully coexist on the quantum regime too.
Crucially, however, objectivity relies on distributed classical information that could conflict with thermalisation. Here, we examine the overlap between thermal and objective states. We find that in general, one cannot exist when the other is present. However, there are certain regimes where thermality and objectivity are more likely to  coexist: in the high temperature limit, at the non-degenerate low temperature limit, and when the environment is large. This is consistent with our experiences that everyday-sized objects can be both thermal and objective.}


\begin{document}
%%%%%%%%%%%%%%%%%%%%%%%%%%%%%%%%%%%%%%%%%%

\section{Introduction}
While fundamental quantum mechanics describes how isolated quantum systems evolve under unitary evolution, realistic quantum systems are open, as they interact with external environments that are typically too large to exactly model. In order to account for large external environments without directly simulating them, the theory of open quantum systems has developed  tools that allow us to study a variety of quantum processes~\mbox{\citep{Heinz-PeterBreuer2007,Rivas2012}}, including decoherence~\citep{Schlosshauer2019} (the loss of phase information to the environment) and dissipation (the loss of energy to the environment)~\citep{dittrich1998quantum}.

The environment, when acting as a heat bath, can lead to the equilibration and thermalisation of quantum
systems~\citep{Yukalov2011,Linden2009,Riera2012,Gogolin2016,Goold2016}.
Meanwhile, in an approach to the quantum-to-classical transition called \emph{Quantum Darwinism}~\citep{Zurek2009,Horodecki2015,Le2019,Cakmak2021,Touil2021}, 
the environment plays a key role in the process of how quantum systems appear classically objective~\citep{Cakmak2021,Touil2021}---whereby classical objective systems have properties that are equivalently independently verifiable by independent observers. In the realm of open quantum systems, whether one process or another occurs depends on multiple factors, including details of the system--environment interactions, initial states, time regimes, averaging, etc.

The (classical) second law of thermodynamics generally states that entropy increases over time. Following this strictly, we may imagine that in the far distant future, the entire universe will reach an equilibrium where entropy can no longer increase: this concept is known as ``heat death'', which can be found in early writings of Bailly, Kelvin, Clausius and von Helmholtz (see references in~\citep{Brush1996}). An alternative, recent, version of heat death would see a universe composed mostly of vacuum and very far separated particles such that no work is done: this is ``cosmological heat death''~\citep{Adams1997}.  There are some caveats to the concept of heat death of the universe: beyond whether or not thermodynamics can be applied at the universal level, it is known that after a sufficiently long time, Poincar\'e recurrences will return the system/universe to its prior states~\citep{Dyson2002}. Furthermore, the discovery of dark energy and the accelerating rate of expansion of the universe~\citep{Riess1998} leads to other theories of the universe's ultimate fate such as the ``big rip''~\citep{Caldwell2003}.

These caveats aside, on more familiar temporal and spatial human scales, both classical and quantum objects can thermalise.  In fact, thermalisation is quite fundamental: in fairly generic conditions, a local subsystem (of a greater state) will likely be close to thermal~\citep{Popescu2006}. We also see that many everyday physical objects have the same approximate temperature as their environment. This thermality appears to contradict with \emph{objectivity}. In Quantum Darwinism, a system state is considered objective if multiple copies of its information exist, which is mathematically expressed as (classical) correlations between the system and its environment~\citep{Le2019,Horodecki2015}. The quintessential example is of the visual information carried in the photon environment. However, information and correlations have an associated \emph{energy}~\citep{Huber2015,Perarnau-Llobet2015}, and naively, this information should not survive under the process of thermalisation. For example, in the model analysed by \citet{Riedel2012}, some level of objectivity emerges at finite time, before equilibration sets in; in the model analysed by \citet{Mirkin2021}, tuning certain parameters produces either objectivity or thermalisation, but not both.

Furthermore, there is a distance-scale difference. Quantum Darwinism requires strong (classical) correlations between two or indeed many more systems, some of which will invariably be very distant from each other---for example, we can view galaxies billions of light years away. In contrast, thermalisation favours realistic settings that have no or rapidly decaying correlations between distant subsystems of the universe. 

In this paper, we investigate this apparent conflict between thermalisation and objectivity and consider whether or not these two can co-exist. To do this, we analyse the overlap between the set of states that are thermal versus the set of states that are objective---if there is no intersection, then there cannot exist any process that produces jointly thermal-objective states. We examine three different sets of thermal states where either: (1) there is system thermalisation, (2) local system and local environment thermalisation, or (3) global system--environment thermalisation. As greater parts of the system-environment become thermal, the overlap between objectivity and thermalisation reduces, often becoming non-existent for many system--environment Hamiltonians. We also find that large environments have better potential to support both thermality and objectivity simultaneously.

This paper is organised as follows. In Section \ref{sec:Objective-states}, we introduce the mathematical structure of objective states, and in Section \ref{sec:Thermal-states}, we introduce thermalised microcanonical states (for finite systems). Then, in Section \ref{sec:Thermal-system-objective-states} we consider the intersection between objective states versus states with a thermal system. In Section \ref{sec:Local-thermal-system-and}, we consider states with a locally thermal system and a locally thermal environment. In Section \ref{sec:Global-thermal-system-environmen}, we consider a globally thermal system-environment state. We discuss and conclude in Section \ref{sec:Conclusion}.

%%%%%%%%%%%%%%%%%%%%%%%%%%%%%%%%%%%%%%%%%%
\section{Objective States\label{sec:Objective-states}}

In our day-to-day experience, we typically perceive the classical world as being ``objective'': objects appear to exist regardless of whether we personally look at them, and the properties of these objects can be agreed upon by multiple observers. More formally, we can describe objective states as satisfying the following:

\begin{Definition}
\textbf{Objectivity}~\citep{Ollivier2004,Zurek2009,Horodecki2015}:
A system state is \emph{objective} if it is (1) simultaneously accessible to many observers (2) who can all determine the state independently without perturbing it and (3) all arrive at the same result.\label{def:Objectivity}
\end{Definition}

\begin{figure}[H]
\includegraphics[width=0.6\columnwidth]{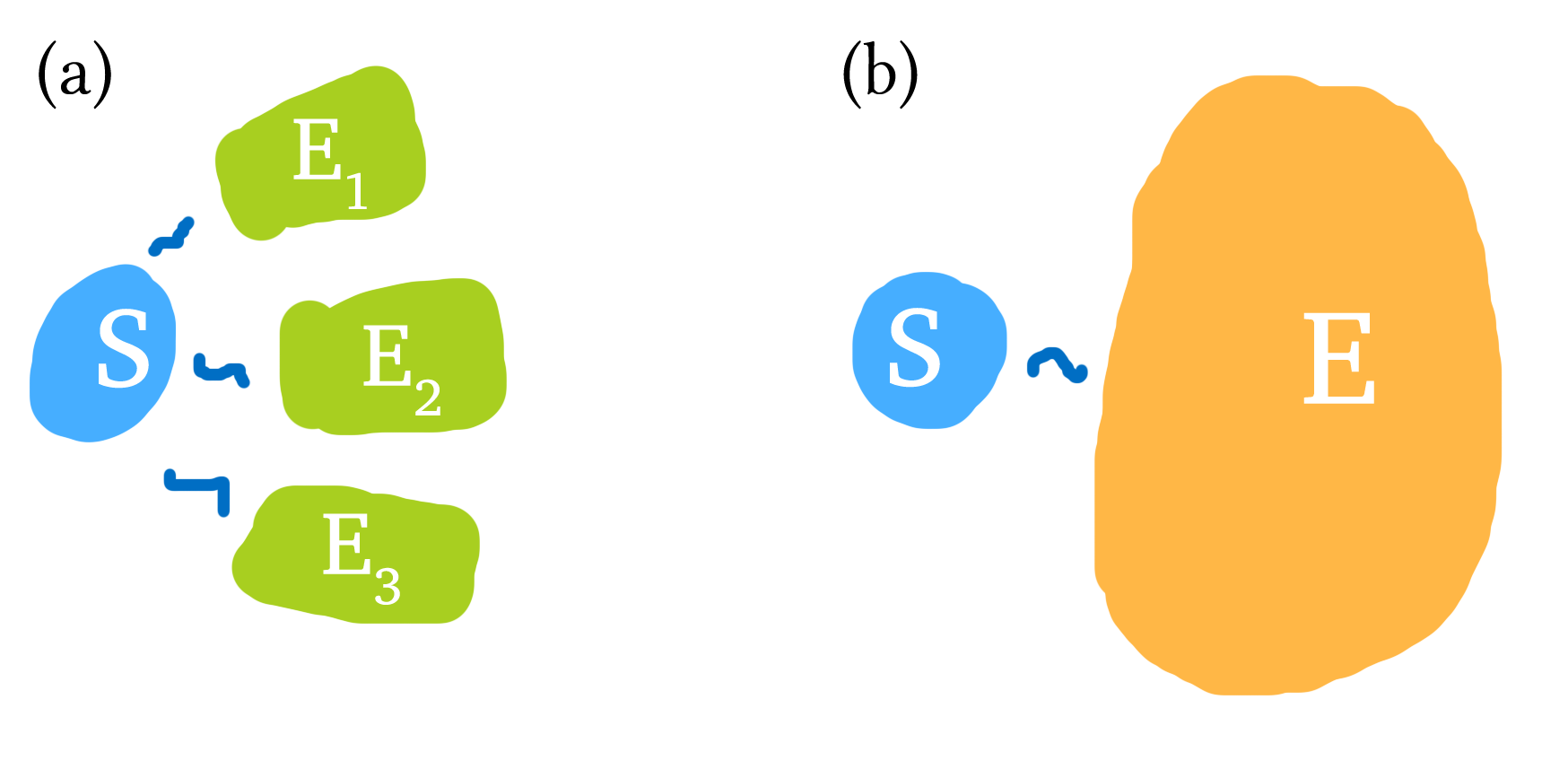}
\caption{(\textbf{a}) Objectivity scenario, where a system interacts with multiple sub-environments, such that those sub-environments contain information about the system.
(\textbf{b}) Thermalisation scenario, where a system interacts with a large heat bath environment and subsequently thermalises to the environment temperature. \label{fig:objective-thermal}}
\end{figure}

The process of emergent objectivity may be described by Quantum Darwinism~\citep{Zurek2003,Zurek2009}: as a system interacts and decoheres due to the surrounding environment, information about the system can spread into the environment. The ``fittest'' information that can be copied tends to record itself in the environment at the expense of other information, thus the name Quantum Darwinism. The paradigmatic example is the photonic environment: multiple photons interact with a physical object and gain information about its physical features, such as position, colour, size, etc. Multiple independent observers can then sample a small part of this photonic environment to find very similar information about the same system state, thus deeming it objective. We depict this in Figure~\ref{fig:objective-thermal}a.

There are a number of frameworks to mathematically describe objective states: in order of increasing restriction one has (Zurek's) Quantum Darwinism~\citep{Zurek2009}, Strong Quantum Darwinism~\citep{Le2019} and Spectrum Broadcast Structure~\citep{Horodecki2015} (and invariant spectrum broadcast structure~\citep{Le2020a}). In this work, we will be focusing primarily on a bipartite system-environment, in which case Strong Quantum Darwinism and Spectrum Broadcast Structure coincide.
In particular, Spectrum Broadcast Structure gives us a clear geometric state structure which is ideal for state analysis.

Objective states with spectrum broadcast structure can all be written in the following form~\citep{Horodecki2015}:
\begin{equation}
\rho_{\mathcal{SE}}=\sum_{i}p_{i}\ket{i}\bra{i}\otimes\bigotimes_{k=1}^{N}\rho_{\mathcal{E}_{k}|i},\quad\rho_{\mathcal{E}_{k}|i}\rho_{\mathcal{E}_{k}|j}=0\,\,\forall i\neq j, \label{eq:SBS_condition}
\end{equation}

\noindent where $\mathcal{E}$ is the accessible environment and $\mathcal{E}_{k}$ are the sub-environments. The conditional states  $\left\{ \rho_{\mathcal{E}_{k}|i}\right\} $ can be used to perfectly distinguish index $i$, where $\left\{ \ket{i}\right\} $ is some diagonal basis of the system and $\left\{ p_{i}\right\} $ its spectrum.
In general, there is no basis dependence in both the system and the environments, and so the overall set of all objective states is non-convex.

\section{Thermal States\label{sec:Thermal-states}}

Systems can exchange energy and heat through interactions with an external environment that functions as a heat bath. Over time, systems can reach thermal equilibrium. Canonically, the thermal state of a quantum system is the Gibbs state \cite{Gogolin2016}. For a given energy/Hamiltonian expectation value, the thermal Gibbs state maximises the von Neumann entropy \cite{Jaynes1957}.

The Gibbs state, which we  denote as $\gamma$, is defined with reference to its Hamiltonian $\hat{H}$ and inverse temperature $\beta=1/k_{B}T$:
\begin{equation}
\gamma=\dfrac{e^{-\beta\hat{H}}}{Z_{\beta,\hat{H}}},
\end{equation}
where $Z_{\beta,\hat{H}}=\tr\left[e^{-\beta\hat{H}}\right]$ is the partition function. 

If the Hamiltonian has the spectral decomposition $\hat{H}=\sum_{i}E_{i}\ket{i}\bra{i}$, then we can write the canonical thermal state as
\begin{equation}
\gamma=\dfrac{\sum_{i}e^{-\beta E_{i}}\ket{i}\bra{i}}{\left(\sum_{j}e^{-\beta E_{j}}\right)}.
\end{equation}

\begin{Remark}
\label{remark:all_states_are_thermal}For any state $\rho$ of full rank, there exists a Hamiltonian $\hat{H}_{\rho}$ and an inverse temperature $\beta_{\rho}$ such that $\rho$ can be considered a thermal state, i.e., we can write $\rho = \frac{1}{Z} \exp[-\beta \hat{H}_{\rho}]$. To see this, suppose the state $\rho$ has the spectral decomposition $\rho=\sum_{i}p_{i}\ket{\psi_{i}}\bra{\psi_{i}}$ ($p_i>0$). Then, consider a Hamiltonian with the same eigenvectors, $\hat{H}_{\rho}=\sum_{i}E_{i}\ket{\psi_{i}}\bra{\psi_{i}}$, with unknown eigenenergies $\left\{ E_{i}\right\} $. We want to find $\left\{ \beta_{\rho},E_{i}\right\} $ such that
\begin{equation}
\dfrac{e^{-\beta_{\rho}E_{i}}}{\left(\sum_{j}e^{-\beta_{\rho}E_{j}}\right)}\overset{!}{=}p_{i},\qquad\forall\,i.
\end{equation}

As there is an extra variable in the set $\left\{ \beta_{\rho},E_{i}\right\} $ compared to the number of conditions, $\left|\left\{ p_{i}\right\} \right|$, this forms an underdetermined set of equations and there can be infinitely many solutions formed by scaling $\beta_{\rho}$ and $E_{i}$ inversely.

\end{Remark}

Objective states are not globally full-rank, but we could add in a very small (non-objective) perturbation to make it full-rank. Then, from this perspective, for any full-rank approximately objective state, we can post-select system-environment Hamiltonians and a temperature at which that objective state is also thermal. In a controlled scenario (e.g., with control of the system Hamiltonian and reservoir engineering~\citep{Schirmer2010}, or in quantum simulators~\citep{Barreiro2011}), it is possible to engineer an approximately objective-and-thermal state by choosing system-environment Hamiltonians based on the objective state itself.

In the rest of this paper, we will be considering the reverse scenario, i.e., \emph{given} some system and environment Hamiltonians and inverse temperature $\beta$, can the subsequent thermal state(s) also support objectivity? By answering this question, we will better understand whether or not objectivity and thermalisation can coexist, and what conditions would allow any coexistence.

In order to answer whether or not there is any overlap between thermalisation and objectivity, we consider the precise state structure. If there is no state overlap, then both properties cannot exist simultaneously, in which case there cannot be any dynamics that produces a non-existent state. More generally, if the two set of states are sufficiently close, then perhaps a compromise is possible.

We will be examining three different types of thermal states:
\begin{enumerate}
\item States with system-only thermalisation. This reflects many applications and research where the system thermality is key, and the environment is assumed inaccessible, or when we have multiple environment baths of different temperatures that are independent and serve different functions.
\item States with local-system thermalisation and with local-environment thermalisation. This corresponds to the common move to describe a system and the environment as being thermal relative to the local Hamiltonians. This situation typically assumes that either the interaction is removed by the time thermality happens, or that the interaction Hamiltonian commutes with all local Hamiltonians, or that the interaction is weak.
\item Global system--environment thermalisation. This is particularly important when there are continued non-trivial, non-commuting interactions between the system and environment.
\end{enumerate}

Examining thermal \emph{states} rather than some time-averaged or instantaneous values of observables means that we are considering thermalisation in a strong sense (or that we have assumed that averaging has already been done). The results are also therefore suitable for more static applications of thermal states, e.g., resource theories.

In order to find the overlap between objective and thermal states, our main method is to start with objective states and successively restrict them to satisfy thermality. As thermal states are full-rank, we will be restricting to objective states where the reduced system and environment states are also full-rank.

Note that if the local system state thermalises, e.g., relative to its energy eigenbasis, then it can also be said to have decohered (relative to that energy eigenbasis). However, whether or not objectivity---an extension of decoherence---arises depends on whether the system thermal information can be encoded in the environment.

\section{Objective States with Thermal System\label{sec:Thermal-system-objective-states}}

In this section, we describe the system--environment states that are both objective and have a locally thermal system (and no requirements on the environment thermality or lack thereof).

Consider the situation where a system with self Hamiltonian $\hat{H}_{\mathcal{S}}$ is put in thermal contact with a bath with some temperature $T_{B}$, is left to thermalise, and then de-coupled from the bath. Writing the system Hamiltonian's spectral decomposition as $\hat{H}_{\mathcal{S}}=\sum_{i}E_{i}\ket{i}\bra{i}$ and with fixed inverse temperature $\beta$, the system thermal state is then
\begin{equation}
\gamma_{\mathcal{S}}=\dfrac{e^{-\beta\hat{H}_{\mathcal{S}}}}{Z_{\beta,\hat{H}_{\mathcal{S}}}}=\dfrac{1}{Z_{\beta,\hat{H}_{\mathcal{S}}}}\sum_{i}e^{-E_{i}\beta}\ket{i}\bra{i}.
\end{equation}
\noindent This implies that objective system-environment states with locally thermal system states must have the following form:
\begin{equation}
\dfrac{1}{Z_{\beta,\hat{H}_{\mathcal{S}}}}\sum_{i}e^{-E_{i}\beta}\ket{i}\bra{i}\otimes\rho_{\mathcal{E}|i},\quad\rho_{\mathcal{E}|i}\rho_{\mathcal{E}|j}=0\,\forall i\neq j,\label{eq:obj_therm_sys_only}
\end{equation}
where the conditional $\rho_{\mathcal{E}|i}$ are perfectly distinguishable. 

As we can immediately see, these objective states describe fixed thermal-state\linebreak information about the system, encoded in the probabilities $\left\{ \dfrac{e^{-E_{i}\beta}}{Z_{\beta,\hat{H}_{\mathcal{S}}}}\right\} _{i}$.
Furthermore, as there are no thermal conditions imposed on the information-carrying environment, the size of set of states satisfying Equation (\ref{eq:obj_therm_sys_only}) is non-empty, as we have freedom to choose any set of mutually distinguishable environment states $\left\{ \rho_{\mathcal{E}|i}\right\} _{i}$. Therefore, objectivity and thermalisation overlap: both can occur at the same time.

The set of exact objective states with thermal system in Equation (\ref{eq:obj_therm_sys_only}) is nowhere dense, as it is a subset of zero-discord states \cite{Ferraro2010}. The set of states in Equation (\ref{eq:obj_therm_sys_only}) is also non-convex in general, though convex subsets can be formed by restricting the conditional subspaces on the environment.

Approximate cases would correspond to imperfect information spreading into the environment and/or imperfect system thermalisation before the information spreading stage. As we have a fairly well-defined set of states {(}Equation (\ref{eq:obj_therm_sys_only}){)}, any distance measure to that set can be used to describe approximately objective-with-thermal-system states,~e.g.,
\begin{equation}
\left[\mathsf{T_{S}O}\right]_{\delta}=\left\{ \rho\middle|\min_{\rho_{obj,th}\in\mathsf{T_{S}O}}\left\Vert \rho-\rho_{obj,th}\right\Vert _{1}\leq\delta\right\} ,
\end{equation}
where $\mathsf{T_{S}O}$ (thermal-system objective) denotes the set of
states satisfying Equation (\ref{eq:obj_therm_sys_only}), and $\left\Vert \cdot\right\Vert _{1}$ is the trace norm. The convex hull of objective-with-thermal-system states are simply zero-discord states with a local thermal system:
\begin{equation}
\left\{ \dfrac{1}{Z_{\beta,\hat{H}_{\mathcal{S}}}}\sum_{i}e^{-E_{i}\beta}\ket{i}\bra{i}\otimes\rho_{\mathcal{E}|i}\middle|\rho_{\mathcal{E}|i}\in\mathcal{H}_{\mathcal{E}}\right\} ,\label{eq:therm_sys_only_conv}
\end{equation}
 i.e., there are no longer any restrictions on the conditional environment states $\rho_{\mathcal{E}|i}$.

\subsection*{Creating Objective States with  Thermal Systems}

A two-step process that produces objective-with-thermal-system states is first system thermalisation followed by information broadcasting. Physically, this can occur if the system was first thermalised using one bath, and then we had a fresh environment interact with the system with intent to gain information. As environments in low-entropy state $\ket{0}$ are typically better for quantum Darwinism~\citep{Giorgi2015,Zwolak2009,Zwolak2010,Balaneskovic2015,Balaneskovic2016}, this second `information-storing' environment could be a very cold bath with states close to the ground state.

The point channel can produce perfectly thermalised states:
\begin{equation}
\Phi_{S,th}\left(\cdot\right)=\tr\left[\cdot\right]\gamma_{\mathcal{S}}.
\end{equation}

One simple method to broadcast information from system to environment  is to start with the information-carrying environment in state $\ket{0}$ (e.g., zero temperature bath). Then, controlled-NOT (CNOT) operations with control system to each individual environment will perfectly broadcast the system information~\citep{Balaneskovic2015,Touil2021}:
\begin{equation}
\Phi_{\text{CNOT}}^{\mathcal{E}_{k}}\left(\rho_{\mathcal{SE}_{k}}\right)  =U_{\text{CNOT}}^{\mathcal{SE}_{k}}\rho_{\mathcal{SE}_{k}}U_{\text{CNOT}}^{\mathcal{SE}_{k}\dagger},
\end{equation}

\noindent where $U_{\text{CNOT}}^{\mathcal{SE}_{k}}$ is the CNOT gate between system $\mathcal{S}$ and environment $\mathcal{E}_{k}$. 

In general, quantum channels that can create the exact objective-with-thermal-system states from Equation~(\ref{eq:obj_therm_sys_only}) are point channels which thermalise the system combined with information broadcasting channels:
\begin{equation}
\Phi_{\mathsf{T_{S}O}}\left(\rho_{\mathcal{SE}}\right)=\dfrac{1}{Z_{\beta,\hat{H}_{\mathcal{S}}}}\sum_{i}e^{-E_{i}\beta}\ket{i}\bra{i}\otimes\Phi_{\mathcal{E}|i}\left(\rho_{\mathcal{SE}}\right),
\end{equation}

\noindent where $\left\{ \Phi_{\mathcal{E}|i}:\mathcal{H}_{\mathcal{S}}\otimes\mathcal{\hat{H}_{\mathcal{E}}}\rightarrow\mathcal{\hat{H}_{\mathcal{E}}}\right\} _{i}$ are channels on the environment such that the output states for different $i$ are orthogonal.

This process can be performed on a quantum simulator by dividing the available qubits into `system', `thermal environment' and `information-carrying environment', and enacting the suitable gate operations \cite{chisholm2021witnessing}.

We can also consider partial thermalisation channels $\Lambda_{p-th}$, such that \emph{repeated application} brings the system closer and closer to thermalisation, i.e.,
\begin{equation}
\Lambda_{p-th}\circ\cdots\circ\Lambda_{p-th}\left(\cdot\right)\rightarrow\gamma_{\mathcal{S}}.
\end{equation}

\noindent If the system is a qubit, then we can, without loss of generality, consider the system qubit Hamiltonian to be $H=\sigma_{z}/2$. One channel which, through repeated application, will lead to the system thermalising is the generalised amplitude damping channel~\citep{nielsenandchuang2010}
\begin{equation}
\rho\left(t\right)=\Phi_{t}^{T}\left(\rho_{0}\right)=\sum_{i=1}^{4}E_{i}\rho_{0}E_{i}^{*},
\end{equation}
with Kraus operators
\begin{align}
E_{1}  &=\sqrt{p}\begin{bmatrix}1 & 0\\
0 & \sqrt{\eta}
\end{bmatrix},\,
&E_{2}  &=\sqrt{p}\begin{bmatrix}0 & \sqrt{1-\eta}\\
0 & 0
\end{bmatrix}, \\
E_{3} &=\sqrt{1-p}\begin{bmatrix}\sqrt{\eta} & 0\\
0 & 1
\end{bmatrix},\,
&E_{4} &=\sqrt{1-p}\begin{bmatrix}0 & 0\\
\sqrt{1-\eta} & 0
\end{bmatrix},
\end{align}

\noindent where $p\in\left[0,1\right]$ depends on the temperature of the environment,
and $\eta_{t}=1-e^{-\left(1+2\bar{N}\right)t}$, where $\bar{N}=\dfrac{1}{e^{1/T}-1}$
is the boson occupation number. The equivalent Bloch sphere representation 
is~\citep{nielsenandchuang2010,Fujiwara2004}
\begin{equation}
\begin{bmatrix}x^{\prime}\\
y^{\prime}\\
z^{\prime}
\end{bmatrix}=\begin{bmatrix}\sqrt{\eta}\\
 & \sqrt{\eta}\\
 &  & \sqrt{\eta}
\end{bmatrix}\begin{bmatrix}x\\
y\\
z
\end{bmatrix}+\begin{bmatrix}0\\
0\\
\left(2p-1\right)\left(1-\eta\right)
\end{bmatrix},
\end{equation}
with stationary state
\begin{equation}
\sigma_{\infty}=\begin{bmatrix}p & 0\\
0 & 1-p
\end{bmatrix},
\end{equation}
where $x=\tr[\sigma_x \rho_0 ]$, $y=\tr[\sigma_y \rho_0 ]$ and $z=\tr[\sigma_z \rho_0 ]$.

More generally, the following channel, in the Bloch sphere representation, will partially thermalise the system:
\begin{equation}
\Lambda_{p-th}\left(\vec{r}\right)=A\vec{r}+\left(1-A\right)\vec{t}_{S},
\end{equation}
where $\vec{t}_{S}$ is the Bloch vector of the system thermal state $\gamma_{\mathcal{S}}$, $\left\Vert A\right\Vert <1$ (under matrix norm) and $\left\Vert A\vec{r}+\left(1-A\right)\vec{t}_{S}\right\Vert_2 \leq1$ for all  $\left\Vert\vec{r}\right\Vert_2\leq1$ (under Euclidean norm). Under repeated application, the state will converge towards the Bloch vector $\vec{t}_{S}$, i.e., to the thermal state.

Aside from the specific model-dependent methods to produce objective-thermal states, it is possible to produce a quantum circuit that will prepare that state \cite{Plesch2011,Araujo2021}. Alternatively, one could also construct a Lindblad generator $\mathcal{L}$  (with an unobserved environment) that simulates a chosen quantum channel (in the infinite time limit) \cite{Albert2016}. In general, the specific timescales will depend on the situation and also the size of the ``unobserved'' environment in comparison with the system and observed environment \cite{Short2012,Brandao2012,Cramer2012,Hutter2013}.

\section{Objective States with Thermal System and Thermal Environment\label{sec:Local-thermal-system-and}}

Thermal environments play a large role in thermodynamics and open quantum systems. In this section, we suppose that both the system and the environment are locally~thermal. 

As in the previous section, we take the system local Hamiltonian to have some general spectral decomposition $\hat{H}_{\mathcal{S}}=\sum_{i}E_{i}\ket{i}\bra{i}$. Suppose that the environment's self-Hamiltonian has this spectral decomposition: $\hat{H}_{\mathcal{E}}=\sum_{k}h_{k}\ket{\psi_{k}}\bra{\psi_{k}}$. This leads to the environment thermal~state 
\begin{equation}
\gamma_{\mathcal{E}}=\dfrac{e^{-\beta\hat{H}_{\mathcal{E}}}}{Z_{\beta,\hat{H}_{\mathcal{E}}}}.
\end{equation}

States that are locally thermal in the system and the environment can be written generally as
\begin{equation}
\rho_{\mathcal{SE}}=\gamma_{\mathcal{S}}\otimes\gamma_{\mathcal{E}}+\chi_{\mathcal{SE}},
\end{equation}

\noindent where $\chi_{\mathcal{SE}}$ is a correlation matrix where $\tr_{\mathcal{S}}\chi_{\mathcal{SE}} = 0$ and $\tr_{\mathcal{E}} \chi_{\mathcal{SE}} = 0$ \cite{Cheong2009}. Our aim is to determine whether this correlation matrix can hold objective correlations.

If the system and environment have \emph{pure} thermal states, then the combined system--environment thermal state $\ket{\gamma_{\mathcal{S}}}\bra{\gamma_{\mathcal{S}}}\otimes\ket{\gamma_{\mathcal{E}}}\bra{\gamma_{\mathcal{E}}}$ is also trivially objective, because there is only one index on the system that the environment needs to distinguish. This can happen if the system and environment only have one energy level, or if the temperature is zero (or very low) and the system and environment both have non-degenerate ground states.

In general though, the system will not have a pure thermal state. With the added restriction of thermal environments, exact co-existence of states that are  simultaneously objective and thermal becomes difficult to achieve: the thermality of the environment comes in conflict with the strong condition of classical correlations required by objectivity.

\subsection{Equal System and Environment Dimension\label{subsec:If-the-system-env-same-dim}}

In the scenario where the system and the individual environments have the same dimension, an exact thermal and bipartite-objective state can only exist for highly fine-tuned system and environment Hamiltonians, i.e., the energy spacing of both must be the~same.
\begin{Remark}
If the system and individual environments have the same dimension, there exists a joint state that is both locally-thermal and objective only if they have the same thermal eigen-energies, i.e., the system Hamiltonian eigen-energies $\left\{ E_{i}\right\} $ differ from the environment Hamiltonian eigen-energies $\left\{ h_{i}\right\} $ by a constant shift, $E_{i}=h_{i}+c$ $\forall i$. \label{rem:same_dimension}
\end{Remark}
\begin{proof}[Proof of Remark~\ref{rem:same_dimension}]
To see this, consider the objective state structure in Equation~\eqref{eq:SBS_condition} and enforce the requirement of local thermality. As the environment has the same dimension, the conditional environment states of the objective state must be pure, and orthogonal for $i\neq j$, i.e., have form $\rho_{\mathcal{E}_k|i}=\ket{\phi_{i|k}}\bra{\phi_{i|k}}$. This leads to the following state which is objective:
\begin{equation}
\rho_{\mathcal{SE}}  =\sum_{i}p_{i}\ket{i}\bra{i}\bigotimes_k\ket{\phi_{i|k}}\bra{\phi_{i|k}},\,\braket{\phi_{i|k}|\phi_{j|k}}=0\,\forall i\neq j,\,\forall k,
\end{equation}
where $\left\{ \ket{\phi_{i|k}}\right\} $ are the eigenvectors of the individual environments. This objective structure corresponds to invariant spectrum broadcast structure~\citep{Le2020a}, as the environment states are also objective.

Local thermality of the system and environments means that
\begin{equation}
\rho_{\mathcal{S}} =\sum_{i}p_{i}\ket{i}\bra{i}\overset{!}{=}\gamma_{\mathcal{S}} \mbox{ and }
\rho_{\mathcal{E}_k} =\sum_{i}p_{i}\ket{\phi_{i|k}}\bra{\phi_{i|k}}\overset{!}{=}\gamma_{\mathcal{E}_k}.
\end{equation}
In order for this to be true, the eigenvalues of both the system thermal state $\gamma_{\mathcal{S}}$ \emph{and} the environment thermal states $\gamma_{\mathcal{E}_k}$ must be identical and equalling $\left\{ p_{i}\right\} $, i.e.,
\begin{equation}
\dfrac{e^{-\beta E_{i}}}{Z_{\beta,\hat{H}_{\mathcal{S}}}}=\dfrac{e^{-\beta h_{i}}}{Z_{\beta,\hat{H}_{\mathcal{E}_k}}} \qquad \forall i,
\end{equation}
with appropriate labelling of ``$i$'' on the system and the environment.

As the inverse temperature is fixed at some $\beta$, this means that the Hamiltonian eigenenergies of the system and environment must also be the same, $\{E_i\}$ and $\{h_i\}$, respectively, up to a constant shift. That is, the environment eigenenergies are $h_i=E_{i}+c$, thus
\begin{equation}
\dfrac{e^{-\left(E_{i}+c\right)\beta}}{\sum_{j}e^{-\left(E_{j}+c\right)\beta}}=\dfrac{e^{-c\beta}e^{-E_{i}\beta}}{\sum_{j}e^{-c\beta}e^{-E_{j}\beta}}=\dfrac{e^{-E_{i}\beta}}{Z_{\beta,\hat{H}_{\mathcal{S}}}},
\end{equation}
as required.
\end{proof}

Realistically, the scenario of system and environments having identical dimension and equal eigenenergies can occur if both are made out of the same \emph{material}, e.g., they are all photons, all spins, etc. with the same internal and external Hamiltonians up to a constant energy shift.

This shows that randomly independently chosen individual Hamiltonians for the system and the environment, will, in general, \emph{not} support an exact thermal and objective system--environment state. Once a particular system Hamiltonian is chosen, say $\hat{H}_{\mathcal{S}}=\sum_{i}E_{i}\ket{i}\bra{i}$, an exact thermal-objective system-environment state (with identical system and sub-environment dimensions) can only exist if the environment Hamiltonians have form $\hat{H}_{\mathcal{E}_k}=\sum_{i}\left(E_{i}+c_k\right)U_k\ket{i}\bra{i}U_k^{\dagger}$, with freedom in real value energy $c_k$ and unitary rotation $U_k$ that produces various sets of orthogonal eigenvectors, in order to give rise to the exact thermal-objective state:
\begin{equation}
\rho_{\mathcal{SE}}^{obj,th}=\dfrac{1}{Z_{\beta,\hat{H}_{\mathcal{S}}}}\sum_{i}e^{-E_{i}\beta}\ket{i}\bra{i}\otimes\bigotimes_{k=1}^{N}\ket{\phi_{i|k}}\bra{\phi_{i|k}},\label{eq:obj-therm-sys-sub-envs}
\end{equation} 
 where $\ket{\phi_{i|k}}=U_k\ket{i}_{\mathcal{E}_k}$.

\subsubsection{Approximate Thermal-Objective States}

As noted, an exact thermal-objective state can only emerge when the system and environment Hamiltonians have a very particular relationship. More generally, we can look for the existence of a state that is {\em approximately} thermal and objective. 

Suppose we allow a deviation in the environment Hamiltonian from the ideal Hamiltonian, i.e., where $\hat{H}_{\mathcal{E}}=\sum_{i}\left(E_{i}+c+\delta_{i}\right)\ket{\phi_{i}}\bra{\phi_{i}}_{\mathcal{E}}$, where $\left\{ \delta_{i}\right\} _{i}$ are different for at least two $i$'s (we work with one environment for simplicity). In this situation, while the state in Equation~\mbox{(\ref{eq:obj-therm-sys-sub-envs})} is objective, it no longer has local thermal environments. We can measure the minimum distance between the set of thermal states and the set of objective states with the trace norm as follows:
\begin{equation}
D_{\text{obj-thm}}\left(\hat{H}_{\mathcal{S}},\hat{H}_{\mathcal{E}},\beta\right)=\min_{\rho_{obj},\gamma_{\mathcal{SE}}}\left\Vert \rho_{obj}-\gamma_{\mathcal{SE}}\right\Vert_1 ,
\end{equation}
where $\rho_{obj}$ are objective states, and $\gamma_{\mathcal{SE}}=\gamma_{\mathcal{S}}\otimes\gamma_{\mathcal{E}}+\chi_{\mathcal{SE}}$ have locally thermal system and environment and variable  correlation matrix $\chi_{\mathcal{SE}}$. 

Taking the ansatz
\begin{equation}
\rho_{obj}^{*}=\dfrac{1}{Z_{\beta,\hat{H}_{\mathcal{S}}}}\sum_{i}e^{-E_{i}\beta}\ket{i}\bra{i}\otimes\ket{\phi_{i}}\bra{\phi_{i}}_{\mathcal{E}},
\end{equation}
from Equation~(\ref{eq:obj-therm-sys-sub-envs}), the distance of this objective state to the set of locally thermal states can be bounded above:
\begin{align}
&D_{\text{obj-thm}}\left(\hat{H}_{\mathcal{S}},\hat{H}_{\mathcal{E}_{k}},\beta\right) \nonumber \\
& \leq\min_{\chi_{\mathcal{SE}}\text{ traceless}}\left\Vert \begin{array}{l}
\dfrac{1}{Z_{\beta,\hat{H}_{\mathcal{S}}}}\sum_{i}e^{-E_{i}\beta}\ket{i}\bra{i}\otimes\ket{\phi_{i}}\bra{\phi_{i}}_{\mathcal{E}}\\
-\dfrac{1}{Z_{\beta,\hat{H}_{\mathcal{S}}}}\sum_{i}e^{-E_{i}\beta}\ket{i}\bra{i}\otimes\dfrac{1}{Z_{\beta,\hat{H}_{\mathcal{E}}}}\sum_{j}e^{-\left(E_{j}+c+\delta_{j}\right)\beta}\ket{\phi_{j}}\bra{\phi_{j}}_{\mathcal{E}}-\chi_{\mathcal{SE}}
\end{array}\right\Vert_1 .
\end{align}
By picking a sample matrix, 
\begin{equation}
\chi_{\mathcal{SE}}=\dfrac{1}{Z_{\beta,\hat{H}_{\mathcal{S}}}}\sum_{i}e^{-E_{i}\beta}\ket{i}\bra{i}\otimes\ket{\phi_{i}}\bra{\phi_{i}}_{\mathcal{E}}-\gamma_{\mathcal{S}}\otimes\dfrac{1}{Z_{\beta,\hat{H}_{\mathcal{S}}}}\sum_{i}e^{-E_{i}\beta}\ket{\phi_{i}}\bra{\phi_{i}}_{\mathcal{E}},
\end{equation}

\noindent the distance is then bounded as
\begin{align}
&D_{\text{obj-thm}}\left(\hat{H}_{\mathcal{S}},\hat{H}_{\mathcal{E}_{k}},\beta\right)  \nonumber \\
& \leq\left\Vert \begin{array}{l}
\gamma_{\mathcal{S}}\otimes\dfrac{1}{Z_{\beta,\hat{H}_{\mathcal{E}}}}\sum_{j}e^{-\left(E_{j}+c+\delta_{j}\right)\beta}\ket{\phi_{j}}\bra{\phi_{j}}_{\mathcal{E}}-\gamma_{\mathcal{S}}\otimes\dfrac{1}{Z_{\beta,\hat{H}_{\mathcal{S}}}}\sum_{i}e^{-E_{i}\beta}\ket{\phi_{i}}\bra{\phi_{i}}_{\mathcal{E}}\end{array}\right\Vert_1 \\
 & =\left\Vert \begin{array}{l}
\dfrac{1}{Z_{\beta,\hat{H}_{\mathcal{E}}}}\sum_{j}e^{-\left(E_{j}+c+\delta_{j}\right)\beta}\ket{\phi_{j}}\bra{\phi_{j}}_{\mathcal{E}}-\dfrac{1}{Z_{\beta,\hat{H}_{\mathcal{S}}}}\sum_{i}e^{-E_{i}\beta}\ket{\phi_{i}}\bra{\phi_{i}}_{\mathcal{E}}\end{array}\right\Vert_1 \\
 & =\sum_{i}\left|\dfrac{e^{-\left(E_{i}+c+\delta_{i}\right)\beta}}{Z_{\beta,\hat{H}_{\mathcal{E}}}}-\dfrac{e^{-E_{i}\beta}}{Z_{\beta,\hat{H}_{\mathcal{S}}}}\right|.\label{eq:subenv_dist_to_thermality}
\end{align}

\noindent The distance is bounded by the difference between the thermal-state eigenenergies, which here is a nonlinear function of the deviations $\{\delta_i\}$.

\begin{figure}[H]
\includegraphics[width=0.75\columnwidth]{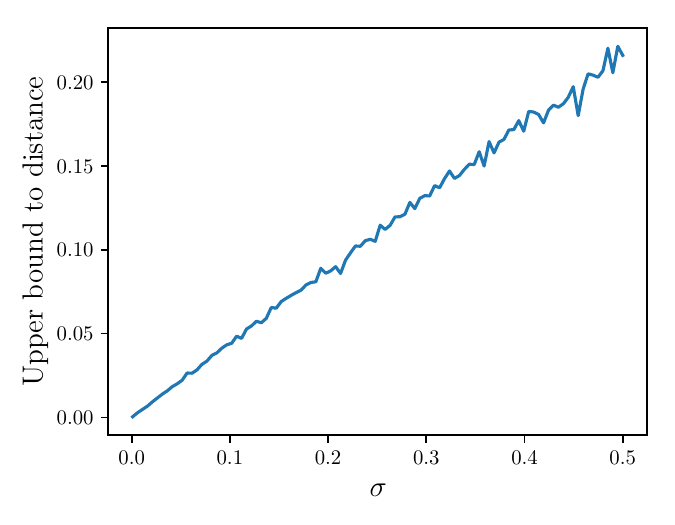}
\caption{Averaged upper bound to the distance (Equation (\ref{eq:subenv_dist_to_thermality})) between the set of objective states vs. the set of thermal states (locally thermal system and environment) versus standard deviation $\sigma$ of the deviations $\delta_{i}$. That is, the environment Hamiltonian is less-than-optimal: for a system Hamiltonian energy distribution $\left\{ E_{i}\right\} $, the environment Hamiltonian energies are $\left\{ E_{i}+\delta_{i}\right\} $, where the deviations are   $\delta_{i}\sim\mathcal{N}\left(0,\sigma\right)$ (normal distribution). The inverse temperature is $\beta=1$, with qubit system and qubit environment. Averaged across 1000 random instances. \label{fig:The-error-for-delta-N}}
\end{figure}

In Figure \ref{fig:The-error-for-delta-N}, we consider if this error is Normal-distributed $\delta_{i}\sim\mathcal{N}\left(0,\sigma\right)$ with mean zero and standard deviation $\sigma$. We see that, on average, increasing the spread $\sigma$ linearly increases the upper bound on the distance measure of Equation~(\ref{eq:subenv_dist_to_thermality}) in the domain considered.

\subsubsection{Employing Macrofractions}

A known technique for improving distinguishability of environments is the use of macrofractions, {i.e.}, grouping multiple subenvironments into a greater environment fragment \cite{Mironowicz2017,Mironowicz2018, Korbicz2017}. By doing this, even if the deviation of the environment Hamiltonian energies from the system Hamiltonian energies is large, we may be able to construct an approximate objective-thermal state.

Consider the distance between the set of objective states and the set of states with locally thermal subsystems similarly as above:
\begin{equation}
D_{\text{obj-thm}}\left(\hat{H}_{\mathcal{S}},\{\hat{H}_{\mathcal{E}_{k}}\}_{k=1}^{N},\beta\right)=\min_{\rho_{obj},\gamma_{\mathcal{SE}}}\left\Vert \rho_{obj}-\gamma_{\mathcal{SE}}\right\Vert_1 ,
\end{equation}
where the following state consists of locally thermal system and environments: $\gamma_{\mathcal{SE}}=\gamma_{\mathcal{S}}\otimes \gamma_{\mathcal{E}_1}\otimes\cdots\otimes\gamma_{\mathcal{E}_N} + \chi_{\mathcal{SE}}$, with correlation matrix $\chi_{\mathcal{SE}}$ such that $\tr_\mathcal{S}[\chi_{\mathcal{SE}}]=0$ and $\tr_{\mathcal{E}_k}[\chi_{\mathcal{SE}}]=0$ for all $k$.

Using $\rho_{obj}^{*}=\dfrac{1}{Z_{\beta,\hat{H}_{\mathcal{S}}}}\sum_{i}e^{-E_{i}\beta}\ket{i}\bra{i}\otimes\bigotimes_{k=1}^{N}\ket{\phi_{i}}\bra{\phi_{i}}_{\mathcal{E}_{k}}$ as an example close-by objective state, and with matrix
\begin{equation}
\chi_{\mathcal{SE}}^{*}=\rho_{obj}^{*}-\gamma_{\mathcal{S}}\otimes\bigotimes_{k=1}^{N}\left(\dfrac{1}{Z_{\beta,\hat{H}_{\mathcal{S}}}}\sum_{i}e^{-E_{i}\beta}\ket{\phi_{i}}\bra{\phi_{i}}_{\mathcal{E}_{k}}\right),
\end{equation}
 the distance is then bounded as\newpage
\begin{align}
&D_{\text{obj-thm}}\left(\hat{H}_{\mathcal{S}},\hat{H}_{\mathcal{E}_{1}},\ldots,\hat{H}_{\mathcal{E}_{N}},\beta\right) \nonumber \\
& \leq\left\Vert \begin{array}{l}
\bigotimes_{k=1}^{N}\left(\dfrac{1}{Z_{\beta,\hat{H}_{\mathcal{E}_{k}}}}\sum_{i}e^{-\left(E_{i}+c+\delta_{i|k}\right)\beta}\ket{\phi_{i}}\bra{\phi_{i}}_{\mathcal{E}_{k}}\right)\\
-\bigotimes_{k=1}^{N}\left(\dfrac{1}{Z_{\beta,\hat{H}_{\mathcal{S}}}}\sum_{i}e^{-E_{i}\beta}\ket{\phi_{i}}\bra{\phi_{i}}_{\mathcal{E}_{k}}\right)
\end{array}\right\Vert_1 \\
 & =\sum_{i_{1},\ldots,i_{N}}\left|\dfrac{e^{-\left(E_{i_{1}}+c+\delta_{i_{1}|1}\right)\beta}}{Z_{\beta,\hat{H}_{\mathcal{E}_{1}}}}+\cdots+\dfrac{e^{-\left(E_{i_{N}}+c+\delta_{i_{N}|N}\right)\beta}}{Z_{\beta,\hat{H}_{\mathcal{E}_{N}}}}-\dfrac{e^{-E_{i_{1}}\beta}\cdots  e^{-E_{i_{N}}\beta}}{\left(Z_{\beta,\hat{H}_{\mathcal{S}}}\right)^{N}}\right|.\label{eq:macrofractions_sub_env_bound}
\end{align}

\begin{figure}[H]
\includegraphics[width=0.75\columnwidth]{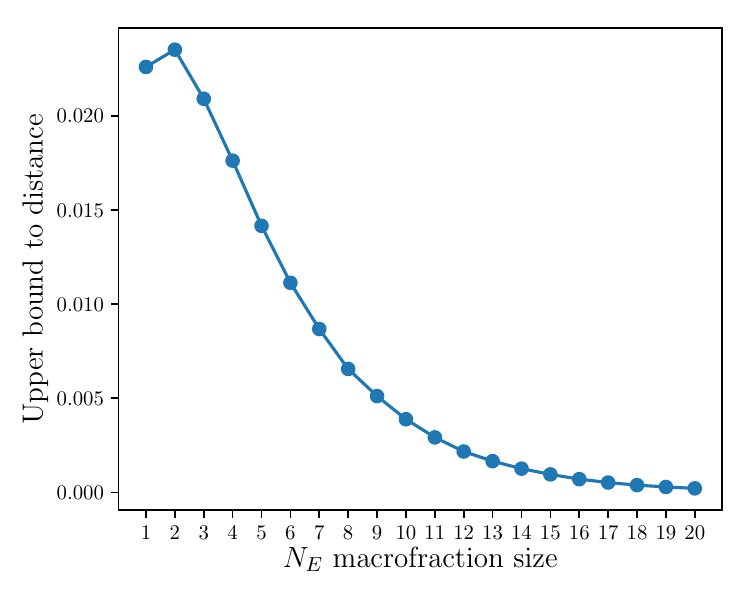}
\caption{Upper bound to the distance (Equation (\ref{eq:macrofractions_sub_env_bound})) between the set of objective states and the set of thermal states (locally thermal system and environment) versus macrofractions of size $N_E$. A macrofraction is collection of environments. Here, the environment Hamiltonians are less-than-optimal, i.e., for a system Hamiltonian energy distribution $\left\{ E_{i}\right\} $, the environment Hamiltonians energies are $\left\{ E_{i}+\delta_{i|k}\right\} $, $k=1,\ldots,N_{E}$, where the ``error'' is $\delta_{i}\sim\mathcal{N}\left(0,\sigma=0.05\right)$ (Normal distribution). The inverse temperature is $\beta=1$, with qubit system and qubit environments. Averaged across 500 random instances. \label{fig:The-error-for-delta-N-1}}
\end{figure}

We plot the behaviour of the bound Equation~\eqref{eq:macrofractions_sub_env_bound} in Figure~\ref{fig:The-error-for-delta-N-1}. As expected, the figure shows that increasing the number of environments included into a macrofraction leads to a decreasing distance between the set of thermal states versus the set of objective states. This is essentially as though we considered increasingly larger environments, which is the focus of the next subsection.

\subsection{Environment Dimension Larger than System Dimension}

In common situations, environments are much larger than the system. Intuitively, larger environment dimensions should give greater flexibility to form approximately objective-thermal states. In this section, we find that the existence of exact thermal-objective states requires very fine tuned system and environment Hamiltonians. However, we will also find that as the dimension of the environment goes up (e.g., towards the classical/thermodynamic limit), there will exist states that are close to both objectivity and local thermality. 

\begin{Theorem}
The distance between the set of objective states and the set of states with locally thermal system and environment goes to zero as the dimension of the environment goes to infinity.\label{thm:obj_therm_as_d_infty}
\end{Theorem}

\begin{proof}[Proof of Theorem~\ref{thm:obj_therm_as_d_infty}]
Consider the distance between these two sets for some given system thermal state $\gamma_{\mathcal{S}}$ and environment thermal state
$\gamma_{\mathcal{E}}$:
\begin{equation}
D_{\text{obj-thm}}\left(\gamma_{\mathcal{S}},\gamma_{\mathcal{E}}\right)=\min_{\stackrel{\rho_{obj}}{\chi_{\mathcal{SE}}}}\left\Vert \rho_{obj} -(\gamma_{\mathcal{S}}\otimes\gamma_{\mathcal{E}}+\chi_{\mathcal{SE}})\right\Vert _{1},\label{eq:theorem1_eq1}
\end{equation}
where $\rho_{obj}$ are objective states and $\chi_{\mathcal{SE}}$ are correlation matrices. Decomposing the system thermal state as $\gamma_{\mathcal{S}}=\sum_{i}p_{i}\ket{i}\bra{i}$,
where $p_{i}=\dfrac{e^{-E_{i}\beta}}{Z_{\beta,\hat{H}_{\mathcal{S}}}}$, we can bound Equation~\eqref{eq:theorem1_eq1} by fixing the local state on the system in the objective states $\rho_{obj}$ as
\begin{equation}
D_{\text{\text{obj-thm}}}\left(\gamma_{\mathcal{S}},\gamma_{\mathcal{E}}\right) \leq\min_{\stackrel{\rho_{\mathcal{E}|i}\perp\rho_{\mathcal{E}|i^{\prime}}}{\chi_{\mathcal{SE}}}}\left\Vert 
\sum_{i}p_{i}\ket{i}\bra{i}\otimes\rho_{\mathcal{E}|i} 
-(\gamma_{\mathcal{S}}\otimes\gamma_{\mathcal{E}}+\chi_{\mathcal{SE}})\right\Vert _{1},\label{eq:theorem1_eq2}
\end{equation}
where $\rho_{\mathcal{E}|i}\perp\rho_{\mathcal{E}|i^{\prime}}$ denotes that the conditional environment states should be perfectly distinguishable as per objectivity. 

By picking a sample matrix $\chi_{\mathcal{SE}}=\sum_{i}p_{i}\ket{i}\bra{i}\otimes\rho_{\mathcal{E}|i}-\gamma_{\mathcal{S}}\otimes\sum_{i}p_{i}\rho_{\mathcal{E}|i}$, the distance Equation~\eqref{eq:theorem1_eq2} is then bounded as
\begin{equation}
D_{\text{\text{obj-thm}}}\left(\gamma_{\mathcal{S}},\gamma_{\mathcal{E}}\right) \leq\min_{\rho_{\mathcal{E}|i}\perp\rho_{\mathcal{E}|i^{\prime}}}\left\Vert \sum_{i}p_{i}\rho_{\mathcal{E}|i}-\gamma_{\mathcal{E}}\right\Vert _{1}.
\end{equation}

Write the environment thermal state as $\gamma_{\mathcal{E}}=\sum_{j}\dfrac{e^{-h_{j}\beta}}{Z_{\beta,\hat{H}_{\mathcal{E}}}}\ket{\psi_{j}}\bra{\psi_{j}}$. Suppose that the states $\rho_{\mathcal{E}|i}$ are diagonal in the same eigenstates as $\left\{ \ket{\psi_{j}}\right\} $, i.e., take $\rho_{\mathcal{E}|i}=\sum_{j}c_{j|i}\ket{\psi_{j}}\bra{\psi_{j}}$, where
$\sum_{j}c_{j|i}=1$, and $c_{j|i}c_{j|i^{\prime}}=0$ for $i\neq i^{\prime}$ for orthogonality. Then,
\begin{equation}
 D_{\text{obj-thm}}\left(\gamma_{\mathcal{S}},\gamma_{\mathcal{E}}\right) \leq\min_{c_{j|i}\text{ orthogonal}}\sum_{j}\left|\sum_{i}p_{i}c_{j|i}-\dfrac{e^{-h_{j}\beta}}{Z_{\beta,\hat{H}_{\mathcal{E}}}}\right|.
\end{equation}
As $c_{j|i}c_{j|i^{\prime}}=0$ (i.e., for orthogonality), we can define disjoint sets $C_{i}$ where $j\in C_{i}$ means $c_{j|i}\neq0$ and $c_{j|i^{\prime}}=0$ if $i\neq i^{\prime}$. We are essentially partitioning the environment eigenvectors $\ket{\psi_j}_\mathcal{E}$ into groups labelled by the \emph{system} eigenvectors $\ket{i}_\mathcal{S}$.

\begin{equation}
 D_{\text{\text{obj-thm}}}\left(\gamma_{\mathcal{S}},\gamma_{\mathcal{E}}\right)\leq\min_{\left\{ C_{i}\right\} \text{disjoint}}\sum_{k=1}^{d_{S}}\sum_{j\in C_{k}}\left|p_{k}c_{j|k}-\dfrac{e^{-h_{j}\beta}}{Z_{\beta,\hat{H}_{\mathcal{E}}}}\right|.
\end{equation}

\noindent Naively, the minimum would occur if $c_{j|i}^{*}=\left(\dfrac{e^{-h_{j}\beta}}{Z_{\beta,\hat{H}_{\mathcal{E}}}}\right)/p_{i}$. However, such $c_{j|i}$ may not lead to a real state, due to lack of normalisation. Instead, we can upperbound this with the candidate $\tilde{c}_{j|i}=\dfrac{c_{j|i}^{*}}{\sum_{k\in C_{i}}c_{k|i}^{*}}$, which \emph{is} normalised. In the optimal case, $\tilde{c}_{j|i}=c_{j|i}^{*}$ and the distance would go to zero. Simplifying, then our candidates are
\begin{equation}
\tilde{c}_{j|i}=\dfrac{e^{-h_{j}\beta}}{\sum_{k\in C_{i}}e^{-h_{k}\beta}},
\end{equation}
 leading to
\begin{equation}
 D_{\text{\text{obj-thm}.}}\left(\gamma_{\mathcal{S}},\gamma_{\mathcal{E}}\right)\leq\min_{\left\{ C_{i}\right\} \text{disjoint}}\sum_{k=1}^{d_{S}}\left|p_{k}-\dfrac{\left(\sum_{j\in C_{k}}e^{-h_{j}\beta}\right)}{Z_{\beta,\hat{H}_{\mathcal{E}}}}\right|.\label{eq:before_the_indices_assignment}
\end{equation}

\noindent Without loss of generality, we can consider the smallest $h_{j}$ to be zero, and therefore $\max_{j}\dfrac{e^{-h_{j}\beta}}{Z_{\beta,\hat{H}_{\mathcal{E}}}}=\dfrac{1}{Z_{\beta,\hat{H}_{\mathcal{E}}}}$.

Consider the following algorithm for picking $j$ indices to include  in $C_{k}$. Every time we include in another index $\tilde{j}$ into $C_k$, the value of $\dfrac{\left(\sum_{j\in C_{k}}e^{-h_{j}\beta}\right)}{Z_{\beta,\hat{H}_{\mathcal{E}}}}$ increases by at most $\dfrac{1}{Z_{\beta,\hat{H}_{\mathcal{E}}}}$. Therefore, a basic procedure is to start with $C_{k}=\left\{ \cdot\right\} $ (empty) and randomly add in $j_{1},j_{2},\ldots$ until we are close to the value of $p_{k}$. We stop adding more $j's$ when $\dfrac{\left(\sum_{j\in C_{k}}e^{-h_{j}\beta}\right)}{Z_{\beta,\hat{H}_{\mathcal{E}}}}$ exceeds the value of $p_{k}$, and can choose to either keep or remove the last $j$ depending on whether its inclusion or exclusion leads to a value closer to $p_{k}$.

Because the maximum step-change is $\dfrac{1}{Z_{\beta,\hat{H}_{\mathcal{E}}}}$, this means that the maximum difference is bounded:
\begin{equation}
\left|p_{k}-\dfrac{\left(\sum_{j\in C_{k}}e^{-h_{j}\beta}\right)}{Z_{\beta,\hat{H}_{\mathcal{E}}}}\right|\leq\dfrac{1}{2}\dfrac{1}{Z_{\beta,\hat{H}_{\mathcal{E}}}}.
\end{equation}
We depict this in Figure~\ref{fig:for_proof_1}.

\begin{figure}[H]
\includegraphics[width=1\columnwidth]{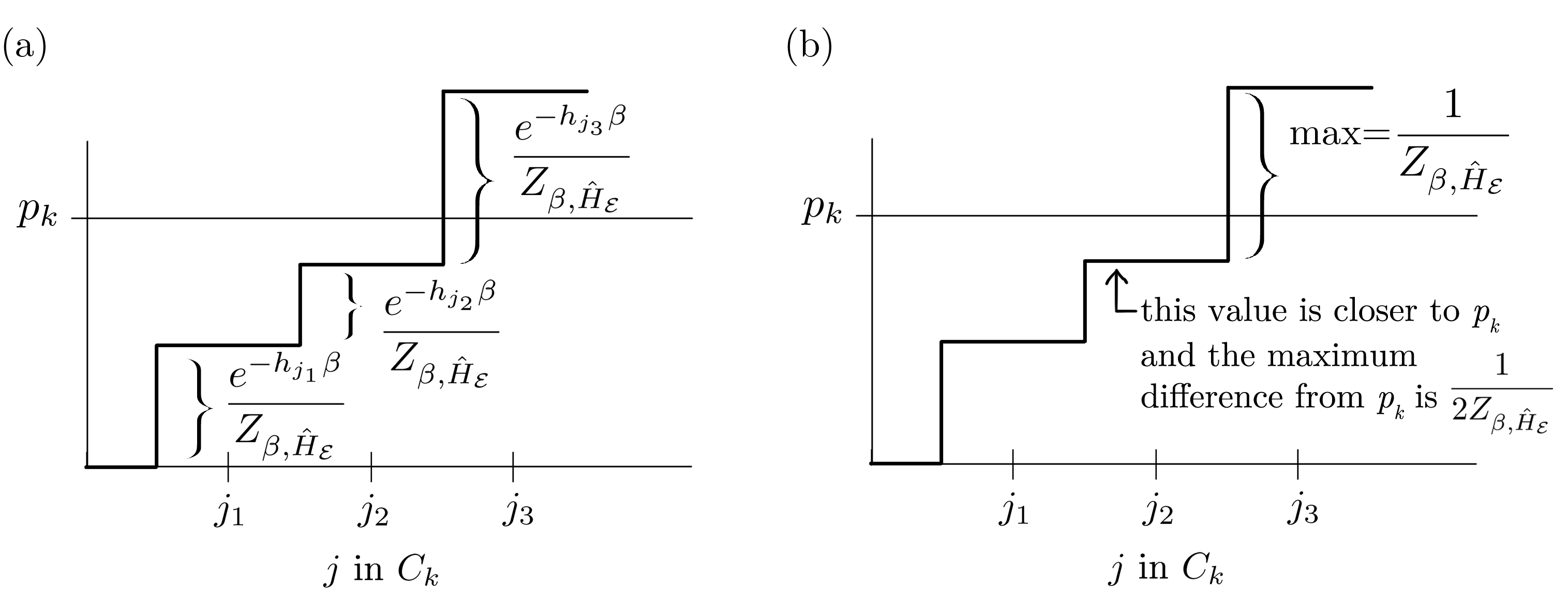}
\caption{{Illustration of} 
part of the proof for Theorem~\ref{thm:obj_therm_as_d_infty}, following after Equation~\eqref{eq:before_the_indices_assignment}. The aim is to assign environment indices $j$ to groups labelled by system indices $k$. As we add in more indices $j$ into $C_{k}$, the sum $\dfrac{\left(\sum_{j\in C_{k}}e^{-h_{j}\beta}\right)}{Z_{\beta,\hat{H}_{\mathcal{E}}}}$ increases. In this example, in (a), we stop adding more indices after $j_{3}$, as $j_{3}$ leads to overshooting the value of $p_{k}$. We either keep the last $j_{3}$ if the sum with $j_{3}$ is closer to $p_{k}$ or we do not include it if the sum is closer to $p_{k}$ without it. In (b), we have decided not to keep the last $j_{3}$ term as the sum is closer to $p_k$ without it.  
\label{fig:for_proof_1}}
\end{figure}

Repeat this for all $p_{k}$. In a random procedure, it may happen that some $C_k$ have been overassigned, leading to many sums which are large $\dfrac{\left(\sum_{j\in C_{k}}e^{-h_{j}\beta}\right)}{Z_{\beta,\hat{H}_{\mathcal{E}}}}>p_{k}$, thus leading to a shortage of indices $j$ left for the remaining $p_{k}$. Therefore, we may have to suboptimally remove earlier indices $\tilde{j}$, leading to a greater maximum difference:
\begin{equation}
\left|p_{k}-\dfrac{\left(\sum_{j\in C_{k}}e^{-h_{j}\beta}\right)}{Z_{\beta,\hat{H}_{\mathcal{E}}}}\right|\leq\dfrac{1}{Z_{\beta,\hat{H}_{\mathcal{E}}}}.
\end{equation}
Therefore,
\begin{equation}
D_{\text{\text{obj-thm}}}\left(\gamma_{\mathcal{S}},\gamma_{\mathcal{E}}\right) \leq\sum_{k=1}^{d_{S}}\dfrac{1}{Z_{\beta,\hat{H}_{\mathcal{E}}}}=\dfrac{d_{S}}{Z_{\beta,\hat{H}_{\mathcal{E}}}}.
\end{equation}

As the dimension of the environment, $d_E$, increases, this leads to more environment Hamiltonian eigenvalues $\left\{ h_{j}\right\} $. This in turn increases the value of the partition function $Z_{\beta,\hat{H}_{\mathcal{E}}}=\sum_{j=1}^{d_E} e^{-h_{j}\beta}\rightarrow\infty$ as $d_E\rightarrow\infty$. Thus, the distance between the set of thermal states and the set of objective states goes to zero: $D_{\text{\text{obj-thm}}}\left(\gamma_{\mathcal{S}},\gamma_{\mathcal{E}}\right)\rightarrow0$ (provided the system dimension remains fixed).
\end{proof}

\subsection{Low Temperature and High Temperature Limits}

Provided that the Hamiltonian of the \emph{system} has a non-degenerate ground state, then in the low temperature limit, the thermal state of the system will be (approximately) pure. At $T=0$, we will have the trivial objective and thermal state $\ket{\psi_{S\text{ground}}}\bra{\psi_{\mathcal{S}\text{ground}}}\otimes\gamma_{\mathcal{E}\text{ ground}}$ (trivially objective in the sense that there is only one index/single piece of information available).

In contrast, in the high temperature limit, the thermal states of the system and environment will approach maximally mixed states. If the dimension of the environment, $d_E$ is the same as the system $d=d_{S}=d_{E}$, then at the infinite temperature limit, the following state satisfies both local thermality and objectivity:
\begin{equation}
\rho_{T\rightarrow\infty}=\dfrac{1}{d}\sum_{i}\ket{i}\bra{i}_{\mathcal{S}}\otimes\ket{\psi_{i}}\bra{\psi_{i}}_{\mathcal{E}},
\end{equation}
along with any other local permutation of indices. This leads to $d!$ different objective-thermal states.

If the dimension of the environment is a \emph{multiple} of the system dimension, then it is also possible to have an exact locally-thermal and objective state: Suppose $d_{E}=Md_{S}$ where $M\in\mathbb{N}$ is a positive integer. Note that the system thermal state at this infinite temperature is $\gamma_{\mathcal{S}}=\sum_{i=1}^{d_{S}}\dfrac{1}{d_{S}}\ket{i}\bra{i}$. The environment thermal state can be written as
\begin{align}
\gamma_{\mathcal{E}} & =\sum_{i=1}^{d_{E}}\dfrac{1}{d_{E}}\ket{\psi_{i}}\bra{\psi_{i}}  =\sum_{i=1}^{Md_{S}}\dfrac{1}{Md_{S}}\ket{\psi_{i}}\bra{\psi_{i}} =\sum_{i=1}^{d_{S}}\dfrac{1}{d_{S}}\rho_{\mathcal{E}|i},\\
 \rho_{\mathcal{E}|i}&\coloneqq\sum_{k=M\left(i-1\right)+1}^{Mi}\dfrac{1}{M}\ket{\psi_{k}}\bra{\psi_{k}}.
\end{align}
Therefore, the joint state
\begin{equation}
\rho_{\mathcal{SE}}=\sum_{i=1}^{d_{S}}\dfrac{1}{d_{S}}\ket{i}\bra{i}\otimes\rho_{\mathcal{E}|i}
\end{equation}
 is both objective \emph{and} satisfies local thermality. We can also see that this state is not unique, \emph{i.e.}, permutations of $\ket{\psi_{k}}$ in each $\rho_{\mathcal{E}|i}$ are possible, thus there is more than one state that is both objective and satisfies local thermality. To be precise, there are $\dfrac{(Md_\mathcal{S})!}{(M!)^{d_\mathcal{S}}}$ such exactly objective-thermal states.

However, in general, the environment dimension is not an exact multiple of the system dimension. Then, in the high temperature limit, there does not exist an \emph{exact} objective-thermal state. We can apply Theorem \ref{thm:obj_therm_as_d_infty} to bound the distance between the set of thermal states (at $T\rightarrow\infty$) and the set of objective states:
\begin{equation}
D_{\text{\text{obj-thm}}}\left(\gamma_{\mathcal{S}},\gamma_{\mathcal{E}}\right)|_{T\rightarrow\infty}\leq\dfrac{d_{S}}{Z_{\beta,\hat{H}_{\mathcal{E}}}}=\dfrac{d_{S}}{d_{E}}.
\end{equation}
 That is, the higher the environment dimension relative to the system dimension, the more likely it is to have a state that is both closely thermal and closely objective.

\section{Objective States That Are Globally Thermal\label{sec:Global-thermal-system-environmen}}

When the system--environment interaction is strong and/or non-commuting with the local Hamiltonians, the thermal state cannot be described by just the local Hamiltonian. Instead, the joint-system environment thermal state is given by the total Hamiltonian, $\hat{H}_{\text{total}}=\hat{H}_{\mathcal{S}}+\hat{H}_{\mathcal{E}}+\hat{H}_{\text{int}}$, where $\hat{H}_{\text{int}}$ is the interaction Hamiltonian:
\begin{equation}
\gamma_{\mathcal{SE}}=\dfrac{e^{-\beta \hat{H}_{\text{total}}}}{Z_{\beta,\hat{H}_{\text{total}}}}.
\end{equation}
 This type of scenario assumes that the system and environment continue to interact for all time, in all the relevant time frames. As there is only one such thermal state for finite systems, we do not have the extra degrees of freedom for forming objective states as we did in the previous two sections. As such, it is highly unlikely that this one global thermal state is also exactly objective. Furthermore, thermal states are full-rank, but exact objective states are not globally full-rank. So at best, there could only an approximately objective-thermal state.

The global thermal state $\gamma_\mathcal{SE}$ will only be approximately objective if the relevant total Hamiltonian structure itself fits a very particular form such that its thermal state is also objective at the appropriate energy scale. The eigenstates of the total Hamiltonian become the eigenstates of the thermal state. Therefore, the Hamiltonians must have a particular \emph{system-environment correlated} eigenstate structure. We give two examples:

\begin{Example}
Consider the Hamiltonian
\begin{equation}
\hat{H}_{\text{total}} =\sum_{i}E_{i}\ket{i}\bra{i}\otimes\ket{\phi_{i}}\bra{\phi_{i}} +\hat{H}_{\text{high-energy}},\label{eq:H_total_1}
\end{equation}
where $\hat{H}_{\text{high-energy}}$ is an orthogonal addition with eigenenergies much higher than the energy scale given by the temperature $T$ and with eigenstates such that $\hat{H}_{\text{total}}$ is full-rank. This produces the following global thermal state that is also approximately objective:

\begin{equation}
\gamma_{\mathcal{SE}} =\dfrac{1}{Z_{\beta,\hat{H}_{\text{total}}}}\sum_{i}e^{-\beta E_{i}}\ket{i}\bra{i}\otimes\ket{\phi_{i}}\bra{\phi_{i}}+\delta_\text{high-energy},
\end{equation}

\noindent where $\delta_\text{high-energy}$ is a perturbative term corresponding to high-energy states.

\end{Example}

\begin{Example}
Consider Hamiltonians of the following form:
\begin{equation}
\hat{H}_{\text{total}} =\sum_{i}E_{i}\ket{i}\bra{i}\otimes\sum_{a}q_{a|i}\ket{\phi_{a}}\bra{\phi_{a}}+\hat{H}_{\text{high-energy}},\label{eq:H_total_2}
\end{equation}
where $q_{a|i}q_{a|j}=0\,\forall i\neq j$,  and where $\hat{H}_{\text{high-energy}}$ is an orthogonal addition with eigenenergies much higher than the energy scale given by the temperature $T$ and with eigenstates such that $\hat{H}_{\text{total}}$ is full-rank. These give rise to a Gibbs thermal state that is also approximately objective:
\begin{align}
\gamma_{\mathcal{SE}} & =\dfrac{1}{Z_{\beta,\hat{H}_{\text{total}}}}\sum_{i,a}e^{-\beta E_{i}q_{a|i}}\ket{i}\bra{i}\otimes\ket{\phi_{a}}\bra{\phi_{a}}  +\delta_\text{high-energy} \\
&=\sum_{i}p_{i}\ket{i}\bra{i}\otimes\sum_{a}c_{a|i}\ket{\phi_{a}}\bra{\phi_{a}} +\delta_\text{high-energy}, \\
p_{i} & \coloneqq\dfrac{\sum_{b}e^{-\beta E_{i}q_{b|i}}}{Z_{\beta,\hat{H}_{\text{total}}}}, \quad c_{a|i} =\dfrac{e^{-\beta E_{i}q_{a|i}}}{\left(\sum_{b}e^{-\beta E_{i}q_{b|i}}\right)}\text{, }c_{a|i}c_{a|j}=0\,\,\forall i\neq j,
\end{align}

\noindent where $\delta_\text{high-energy}$ is a perturbative term corresponding to high-energy states.
\end{Example}

\begin{Remark}
Recall Remark \ref{remark:all_states_are_thermal} where, for any given state of full rank, a Hamiltonian and temperature can be found such that it can be considered thermal. As such, for any full rank approximately objective state, a Hamiltonian and temperature can be found such that it can be considered also thermal.
However, the objective states form a set of measure zero (as discord-free states have zero measure~\citep{Ferraro2010}). Thus, the set of sub-component Hamiltonians directly corresponding to those objective states (up to a mutiplicative coefficient, and not including high-energy terms) is also zero measure.
\end{Remark}

Since objective states are not globally full-rank, there are no objective states that are also exactly globally thermal, and most Hamiltonians will not produce an approximately objective state either. The Hamiltonians that do give rise to (approximately) objective thermal states such as those given in Equations~\eqref{eq:H_total_1} and \eqref{eq:H_total_2}  consist of strong, \emph{constant}, interactions between the system and the environments, which is unrealistic.

\section{Conclusions\label{sec:Conclusion}}

In our everyday experience, there are a number of phenomena which appear natural to us. One of them is \emph{thermalisation}, in which physical objects eventually reach thermal equilibrium with the surrounding environment, e.g., an ice cream melting in hot weather. We also typically take for granted that physical objects are \emph{objective}, i.e., their existence and properties can be agreed upon by many people. On the quantum mechanical level, thermalisation and objectivisation of quantum systems can arise through their interaction with external environments. 

Thermalisation itself is thought to be a generic process and will occur approximately in general scenarios~\citep{Popescu2006}, more so than objectivity~\citep{Brandao2015,Korbicz2017}. In contrast, objectivity requires classical correlations that are more sensitive to the situation, though components of objectivity can occur generically \cite{Brandao2015,Korbicz2017,Knott2018,Colafranceschi2020}.

In general, the set of objective states does not have a preferred basis. Imposing (approximate) thermality can help select a preferred basis on the system and environment, which also leads to a preferred arrangement of classical correlations. If the system local Hamiltonian commutes with the interaction Hamiltonian (among the more straight forward scenarios in which quantum Darwinism has been explored \cite{Tuziemski2015,Mironowicz2017,Tuziemski2019,Roszak2019,Roszak2020}), then the preferred basis of objectivity would coincide with the ``thermal'' basis. The joint analysis of objectivity and thermalisation is further motivated by the fact that we observe everyday classical objects that are both objective and thermal.

In this paper, we examined the intersection of thermalisation and objectivity, especially when a single environment is required to fulfil both roles.  In particular, we examined whether they can exist simultaneously by exploring whether a system-environment state can be both thermal (having the microcanonical Gibbs form) and objective (having state structure that satisfies spectrum broadcast structure).

By sequentially considering whether only the local system is thermal, or the local system and local environment, or the joint system-environment is thermal, we are able to characterise how rare it is for thermality and objectivity to coincide.  This is summarised in Table \ref{tab:summary}. As we increased the thermalisation requirement from local system to global system and environment, the likelihood of an overlapping objectivity-thermal state existing decreases. This shows that in general, thermality and objectivity \emph{are} at odds.

\end{paracol}
\nointerlineskip
\begin{specialtable}[H]
\widetable
\caption{Summary table. $\hat{H}_{\mathcal{S}}$ is the system Hamiltonian, $\hat{H}_{\mathcal{E}}$ is the environment Hamiltonian and $\beta$ is the inverse temperature.\label{tab:summary}}
\setlength{\tabcolsep}{1.5mm}\begin{tabular}{cc}
\toprule
\textbf{Setting}	& \textbf{Coexistence}\\
\midrule
Thermal system only		& Yes, for all $\hat{H}_{\mathcal{S}}$ and $\beta$			\\
Local-thermal system and environment		        & Only for some $\hat{H}_{\mathcal{S}}$, $\hat{H}_{\mathcal{E}}$; an approximate state exists for large environments		\\
Global thermal system and environment        & Only approximate state possible, \emph{extremely} rare and fine-tuned\\
\bottomrule
\end{tabular}
\end{specialtable}
\begin{paracol}{2}
\switchcolumn

By studying the intersection of the sets of thermal and objective states, we can therefore also give a statement about the dynamics that have either objective states or thermal states as their fixed points or as their asymptotic state(s): due to the fine-tuned structure of thermal-objective states, only finely tuned dynamics would produce those states.

Quantum Darwinism can be hindered by numerous factors, such as non-\linebreak Markovianity~\mbox{\cite{Galve2016, Pleasance2017, Le2018, Lorenzo2020, Giorgi2015, Milazzo2019, Lorenzo2020a, Garcia-Perez2020}}, non-ideal environments \cite{Zwolak2009, Zwolak2010}, initial system--environment correlations \cite{Giorgi2015, Balaneskovic2015},  environment--environment interactions \cite{ Riedel2012,Mirkin2021,Giorgi2015,Milazzo2019,Ryan2021}, etc.  It was shown that environment-environment interactions can lead to thermalisation at the detriment of objectivity in ~\cite{Riedel2012,Mirkin2021}, but it is still open whether the other factors would lead to similar behaviour.

Based on these results, we conclude that if the hypothetical entropic death of the universe is characterised by the global thermalisation of the entire (observable) universe, then it is extremely unlikely for objectivity to remain. This is consistent with our intuition that, at thermalisation (heat death), there should be no work left to be done. In contrast, objectivity implies information about one system in another, which usually contains extractable work
 \cite{Perarnau-Llobet2015}.

That said, there are (very) rare situations where a global thermal state can still support objective correlations, at least theoretically. \emph{If} objectivity and information does remain, then this implies that there are highly nonlocal, strong interactions, as such giving rise to Hamiltonians like in Equation~\eqref{eq:H_total_2}, which are required to maintain correlations in the global thermal state. While this is unrealistic that the entire universe can have such strong interactions, it may be possible for smaller parts of the universe to maintain interactions and thus have subcomponents that are objective.

Another possibility is that the system alone thermalises on the short time scale, while on more intermediate timescales the system and (information-carrying) environment locally thermalises. Meanwhile, perhaps only at long time scales does the global system-environment thermalise, achieving an ultimate ``heat death''. We found that objectivity is more likely to be able to coexist with thermality in the first two situations. This suggests that objectivity can survive in the short and intermediate timescales, before fading away at the long timescale.

The following narrative feels intuitive: e.g., decoherence occurs first as a loss of phase information, followed by the classical information spread that characterises objectivity; the classical information fades, followed by thermalisation in which all information is lost (aside from select information such as temperature) \cite{Riedel2012}. Whether this is `common' remains an open question.

%%%%%%%%%%%%%%%%%%%%%%%%%%%%%%%%%%%%%%%%%%
\vspace{6pt} 

%%%%%%%%%%%%%%%%%%%%%%%%%%%%%%%%%%%%%%%%%%
\authorcontributions{Conceptualisation, T.P.L.; methodology, TPL; formal analysis, T.P.L., A.W. and G.A.; writing---original draft preparation, T.P.L.; writing---review and editing, T.P.L., A.W. and G.A.; visualisation, T.P.L. All authors have read and agreed to the published version of the manuscript.}

\funding{TPL acknowledges financial support from the UKRI Engineering and Physical Sciences Research Council (EPSRC) under the Doctoral Prize Award (Grant No.~EP/T517902/1) hosted by the University of Nottingham. GA acknowledges financial support from the Foundational Questions Institute (FQXi) under the Intelligence in the Physical World Programme  (Grant No.~RFP-IPW-1907). AW was supported by the Spanish MINECO (projects FIS2016-86681-P and PID2019-107609GB-I00/AEI/10.13039/501100011033), both with the support of FEDER funds, and by the Generalitat de Catalunya (project 2017-SGR-1127).}

\conflictsofinterest{The authors declare no conflicts of interest.}

%%%%%%%%%%%%%%%%%%%%%%%%%%%%%%%%%%%%%%%%%%
%% Optional
\appendixtitles{no}
\appendixstart
\appendix

%%%%%%%%%%%%%%%%%%%%%%%%%%%%%%%%%%%%%%%%%%
\end{paracol}
\reftitle{References}

\end{document}